\documentstyle[12pt]{article}

\title{{\normalsize{{\hskip 8.5cm} CBPF-NF-078/95}}\\
{\normalsize{{\hskip 8.5cm} BIHEP-TH-95-26}}\\
	Heavy to Light Baryon Transition Form Factors}
\author{Xin-Heng Guo$^{1,3}$, Tao Huang$^{2,3}$ and Zuo-Hong Li$^{3}$\\
       {\small $^{1}$Centro Brasileiro de Pesquisas Fisicas, Rio de Janeiro, Brasil,}\\
	{\small $^{2}$CCAST(World Laboratory)P.O.Box 8730,
        Beijing 100080, P. R. China,}\\
        {\small $^{3}$Institute of High Energy Physics, Academia Sinica,
        Beijing 100039, P. R. China.}}
\date{}
\begin{document}
\maketitle
\begin{abstract}
Recently, Stech found form factor relations for heavy to light transitions
based on two simple dynamical assumptions for spectator partical. In this
work we generalize his approach to the case of baryons and find that for
$\Lambda_{Q}\rightarrow\Lambda$ ($Q$=$b$ or $c$) only one independent form
factor remains in limit $m_{Q}\rightarrow\infty$. Furthermore, combining
with the model of Guo and Kroll we determine both of the two form factors for 
$\Lambda_{Q}\rightarrow\Lambda$ in the heavy quark limit. The results
are applied to $\Lambda_{b}\rightarrow\Lambda + J$/$\psi$ which is not clarified
both theoretically and experimentally. It is found that the branching ratio
of $\Lambda_{b}\rightarrow\Lambda + J/\psi$ is of order 10$^{-5}$.
\end{abstract}
\vspace{0.5cm}
\noindent{\bf PACS numbers:} 12.39.Hg, 12.38.Lg, 12.39.-x, 13.30.-a, 14.20.Lg, 14.20.Mr
\vspace{1.0cm}
\section{Introduction}
\hspace{0.29cm} The heavy quark effective theory (HQET)$^{[1]}$ has been
proven to be a powerful
tool dealing with the physics of hadrons containing a heavy quark, and based 
on it there have been a lot of developments in the study of the heavy flavor 
weak decay. The advantage of HQET in dealing with the weak decays of heavy 
hadrons is that the number of the form factors describing hadronic
matrix elements is reduced. For example, in $\Lambda_b \rightarrow \Lambda_c$
transition all the form factors can be expressed in terms of the Isgur-Wise 
function and a unknown parameter $\bar{\Lambda}$ to order 1/$m_{Q}$ where
$m_{Q}$ is $c$ or $b$ quark mass,
and in the heavy-to-light processes such as $\Lambda_{b}\rightarrow\Lambda$,
the use of HQET in the limit $m_{Q}\rightarrow\infty$ allows one 
to express this transition
in terms of two independent form factors$^{[2]}$. In order to find the relations
for the remaining form factors for the heavy to light transitions, recently, 
Stech$^{[3]}$ proposed a new approach dealing with heavy-to-light transitions 
in the case of mesons where two simple dynamical assumptions for spectator 
particle in the decay process are made. In the present work, we will try
to generalize Stech's work to the baryon cases, i.e., $\Lambda_{b,c}
\rightarrow\Lambda$ transitions. It will be found that this can provide
an additional relation between the form factors $F_{1}$ and $F_{2}$
in the limit $m_{Q}\rightarrow\infty$.\par

The decrease of the form 
factor number can simplify calculations. However, these form factors
contain all soft QCD effects, which are difficult to be calculated from the
fist principle. Therefore, one must resort to some phenomenological
models to calculate them. A very interesting process is $\Lambda_{b}
\rightarrow
\Lambda + J/\psi$. UA1$^{[4]}$ reported the measurement result 
$F(\Lambda_{b})BR(\Lambda_{b}\rightarrow\Lambda + J$/$\psi)=(1.8\times\pm
0.6\pm0.9)\times10^{-3}$ where $F(\Lambda_{b})$ is the fraction of $b$
quark transition into $\Lambda_{b}$, while CDF$^{[5]}$ and OPAL$^{[6]}$
only observed the upper limit $0.5\times10^{-3}$ and $1.1\times10^{-3}$
respectively. On the other hand, theoretically, in refs. [7] and [8] the
authors discussed the process 
$\Lambda_{b}\rightarrow\Lambda + J$/$\psi$ on the basis of HQET and some 
phenomenal considerations. Their results 
for the branching ratios of $\Lambda_{b}\rightarrow\Lambda + J$/$\psi$
are different: 
in [7] $BR(\Lambda_{b}\rightarrow\Lambda + J$/$\psi)$ is of order $10^{-4}$
from quark model calculations, while in [8] it is found that
$BR(\Lambda_{b}\rightarrow\Lambda + J$/$\psi$)=$4\times10^{-5}$ by extracting
form factors at $\omega = 1$ from experiments.  
The up-down asymmetry
parameter $\alpha$ which in fact depends on the ratio between two form
factors in the amplitude of $\Lambda_{b}\rightarrow\Lambda$ 
in heavy quark limit, 
is -0.11 in [7] and 0.25 in [8].
Hence, both theoretically and experimentally, 
the branching ratio of $\Lambda_{b}\rightarrow\Lambda + J$/$\psi$
is very equivocal and to clarify this issue
is necessary.\par

This paper is organized as the following: In sect.2 we recapitulate the Stech's 
approach and then generalize
it to the case of baryons and find a relation between $F_{1}$ and $F_{2}$. To determine $F_{1}$ and $F_{2}$ absolutely, another relation
between $F_{1}$ and $F_{2}$ obtained in the model of Guo and Kroll$^{[9]}$ 
is applied to our case in sect.3.
In sec.4 the branching ratio of $\Lambda_{b}\rightarrow\Lambda + J$/$\psi$ 
in heavy quark limit is obtained.
The last section is devoted to conclusion.\par 

\section{Stech's approach and its generalization to the case of baryons}

\hspace{0.29cm} In this section, we briefly review Stech's
approach$^{[3]}$ to deal with the heavy-to-light transitions in the meson case.
The key point of the Stech's method is the following two 
dynamical assumptions:\par

\hspace{0.4cm} i). In the rest frame of a meson the off-shell energy 
of a constituent
quark is close to its constituent mass independent or little dependent of its
space momentum.\par

\hspace{0.4cm} ii). In the first stage of the weak transition, i.e., before 
final 
hadronization, the spectator quark retains its original momentum and spin.\par
From assumption i), one can think that the off-shell energy
of the spectator quark $\epsilon_{sp}$ in the rest frame of 
a meson is remarkably smaller
than b-quark mass $m_{b}$. Using these two assumptions and 
some Lorentz transition
relations between the initial and final state rest frames, Stech arrives at
three conclusions: (a), In the first stage of the weak transition, the energy
carried by the spectator quark is approximately equal to that of the spectator
in the 
rest frame of the final state particle even for energetic transition, i.e., 
the spectator
doesn't pick up a large energy fraction; (b), In the process
of weak transition, even with large energy release, the relevant b-quark space 
momenta are much smaller than b-quark mass and of the order of confinement
scale. (c), In the first stage of the weak decay the generated $u$ or $c$-quark
carries energy and longitudinal momentum of the final particle, apart from
correction of order $\epsilon_{sp}^{F}$/$E_{F}$ where $\epsilon_{sp}^{F}$ 
and $E_{F}$ are the energy of spectator quark
in final state rest frame and the energy of final particle in the initial 
meson rest frame, respectively.\par

Making use of these conclusions and taking a reasonable assumption
into account that the average of the transverse momentum squared of the 
b-quark ${(\vec{q}_{b{\bot}})}^{2}$ is very
small compared to $E_{F}^{2}$, one can find that the 
transition matrix element
of the weak current corresponding to $b\rightarrow{c}$ or $u$ is 
proportional to the c-number
matrix element T$^{\mu}$
	
\begin{equation}
T^{\mu} = [\bar{U}_{u,c}^{s'}(\vec{p}_{F},m_{u,c})\gamma^{\mu}(1-\gamma_{5})
 {U}_{b}^{s}(\vec{0},m_{b})]L_{s',s},
\end{equation}								
\noindent where $m_{i}$ $(i=u, c, b)$ is the corresponding 
current mass of quark, the
b-quark space momentum  
in the Dirac spinor of the b-quark has been neglected 
due to the conclusion (b). The $L_{s',s}$ are the elements 
of a $2\times2$ spin unit 
matrix. $L=I$ if B decays to a pseudoscalar meson and 
$L$=$\vec{\sigma}\cdot\vec{e}$ if B decays to a vector particle polarized in 
$\vec{e}$ direction. 
A comparison of (1)
with the conventional form factor decomposition$^{[10]}$ gives some form factor
relations$^{[3]}$ in the heavy-to-light transitions. For 
example, for B to pseudoscalar meson via transition $b\rightarrow{u}$ one can 
take $m_{u}$=0 and find

\begin{equation}
F_{1}(q^{2},m_{F}) = R_{u,c}^{B}(q^{2}, m_{F}),
\end{equation} 

\begin{equation}
F_{0}(q^{2}, m_{F}) = ( 1  - \frac{q^{2}}{m_{B}^{2} 
  -m_{F}^{2}} )R_{u,c}^{B}(q^{2}, m_{F}),
\end{equation} 

\noindent where $R_{u,c}^{B}(q^{2}, m_{F})$ is an unknown 
universal function depending not only 
on $q^{2}$ but also on $m_{F}$ and the flavor of outgoing
quarks (and on $m_{B}$).\par

Making a comparison between the heavy-to-light and the heavy-to-heavy form 
factors, we can see
that the main difference between them is that the 
heavy-to-light transition form factors depend on $m_{F}$ due to the 
lack of heavy quark symmetry, while the 
heavy-to-heavy form factors have nothing to do 
with the final state particle.\par

Generally the theoretical predictions$^{[3]}$ from Stech's approach are in
good agreement with experimental data.\par   

In the following, we try to generalize it to the case of baryons.
It is well known that a baryon containing a heavy quark, for 
example, $\Lambda_{b}$,
can be effectively considered as a bound state of a b-quark and a scalar
diquark $S[ud]$ with $[ud]$ quantum numbers. In the transition
$\Lambda_{b,c}\rightarrow\Lambda$, $b$ (or $c$)quark decays into $s$ quark
and the other part $[ud]$ behaves as spectator. This decay picture is almost
same as that of meson case. The only difference is that the spectator 
quark in meson case is replaced by a diquark. The mass of the $S[ud]$ diquark
is about several hundred Mev. In the light of this picture, the two basic 
dynamical assumptions made by Stech can be generalized to the baryon case.
It is straightforward to see that the three conclusions mentioned before
are still valid now.\par
 
$\Lambda_{b}$ and $\Lambda$ can be  
represented by Dirac spinor $U(v)$ and $U(P_{\Lambda})$ 
respectively where $m_{b}v$ and $P_{\Lambda}$ are four momentum of
$\Lambda_{b}$ and $\Lambda$.
In the limit $m_{b}\rightarrow\infty$, the 
matrix element of $\Lambda_{b}\rightarrow\Lambda$ can be written 
as$^{[2]}$

\begin{equation}
<\Lambda(P_\Lambda)\mid\bar{s}\gamma_{\mu}(1-\gamma_{5})b\mid\Lambda_{b}(v)>
 = \bar{U}_{\Lambda}(P_{\Lambda})[F_{1}(v\cdot{P_{\Lambda}})+\rlap/v{F_{2}(v\cdot{P_{\Lambda}})}]
   \gamma_{\mu}(1-\gamma_{5})U_{\Lambda_{b}}(v).
\end{equation}

\noindent According the generalized Stech's approach in the baryon case, 
this matrix element should
be proportional to the following C-number matrix element

\begin{equation}
T_{\Lambda_{b}\rightarrow\Lambda}^{\mu} = \bar{U}_{s}(\vec{P}_{\Lambda},m_{s})
  \gamma_{\mu}(1-\gamma_{5})U_{b}(\vec{0},m_{b})
\end{equation}

\noindent A comparison of (4) and (5) gives

\begin{equation}
F_{1} \sim \sqrt{\frac{(E_{\Lambda} + m_{s})m_{\Lambda}}
  {(E_{\Lambda} + m_{\Lambda})m_{s}}} \frac{2E_{\Lambda}+m_{\Lambda}+m_{s}}
  {2(E_{\Lambda}+m_{s})},
\end{equation}

\begin{equation}
F_{2} \sim \sqrt{\frac{(E_{\Lambda} + m_{s})m_{\Lambda}}
  {(E_{\Lambda} + m_{\Lambda})m_{s}}} \frac{m_{s}-m_{\Lambda}}
  {2(E_{\Lambda}+m_{s})},
\end{equation}

\noindent where $m_{s}$ is current mass of the s-quark 
and $E_{\Lambda}$ is energy of $\Lambda$
in the rest frame of $\Lambda_{b}$. For 
the heavy-to-heavy transition, for instance, 
$\Lambda_{b}\rightarrow\Lambda_{c}$, $m_{c}\simeq{m_{\Lambda_{c}}}$ in heavy quark limit, hence $F_{2}$=0 and $F_{1}$ is 
the only form factor, which is in fact the Isgur-Wise function. This is 
consistent with HQET.
From (6) and (7), we arrive at the form factor ratio

\begin{equation}
 \frac{F_{2}}{F_{1}} = \frac{m_{S}-m_{\Lambda}}
  {2(E_{\Lambda}+m_{S}+m_{\Lambda})}.
\end{equation}
								
\noindent In the rest frame of $\Lambda_{b}$, E$_{\Lambda}$ can be 
represented by the invariant momentum 
transfer $q^{2}(=(P_{\Lambda_{b}} - P_{\Lambda})^{2})$

\begin{equation}
E_{\Lambda} = \frac{1}{2m_{\Lambda_{b}}} (m_{\Lambda_{b}}^{2}
 + m_{\Lambda}^{2} - q^{2}).
\end{equation}	

\noindent Taking $m_{s}$=0.15 GeV, $m_{\Lambda}$=1.116GeV 
and $m_{\Lambda_{b}}$=5.64GeV,
one finds that E$_{\Lambda}$ ranges
from 1.116GeV to 2.93GeV and $F_{2}$/$F_{1}$ varies from -0.28 to -0.14, with 
the q$^2$
from $q_{max}^{2}=(m_{\Lambda_{b}} - m_{\Lambda})^{2}$ to $q^{2}$ = 0. 
Similarly , $F_{2}/F_{1}$ for $\Lambda_{c}\rightarrow\Lambda$ changes from
-0.28 to -0.24. This is in good agreement with experimental value
($-0.25{\pm}0.14{\pm}0.08$) measured recently by CLEO$^{[11]}$.\par    

\section{Overlap Integral for $\Lambda_{b,c}\rightarrow\Lambda$ Form Factors}

\hspace{0.29cm} In last section eq. (8) provides a relation between $F_{1}$ and
$F_{2}$. To determine them absolutely , we prepare to use model adopted by 
Guo and Kroll$^{[8]}$ where they worked in the infinite momentum frame (IFM)
which is arrived at by boosting along the 3-direction
(with $P\rightarrow\infty$ )
from a frame with opposite velocities: 
${P}_{\Lambda_{b}}^{\mu}$
=$m_{\Lambda_{b}}(\sqrt{1+v^{2}/4},-v/2,0,0)$; 
${P}_{\Lambda}^{\mu}$
=$m_{\Lambda}(\sqrt{1+v^{2}/4},v/2,0,0)$, 
the IFM momenta of $\Lambda_{b}$ and $\Lambda$ read, respectively,

\begin{equation}
P_{\Lambda_{b}}^{\mu} = P(1+(1+v^{2}/4)m_{\Lambda_{b}}^{2}/(2P^{2}), 
   {-m_{\Lambda_{b}}{v}}/(2P), 0, 1),
\end{equation}

\begin{equation}
P_{\Lambda}^{\mu} = Pm_{\Lambda}/m_{\Lambda_{b}}(1+(1+v^{2}/4)
    m_{\Lambda_{b}}^{2}/(2P^{2}), 
    {m_{\Lambda_{b}}\cdot{v}}/(2P), 0, 1).
\end{equation}

\noindent A heavy baryon $(\Lambda_{b}$ or $\Lambda_{c})$ is regarded as a 
relativistic bound state of a heavy
Q ($b$ or $c$) of mass $m_{Q}$ and a scalar diquark $S[ud]$,

\begin{equation}
\mid\Lambda_{Q}(\vec{P},\lambda> = \sqrt{\frac{m_{Q}}{2m_{\Lambda_{Q}}}}
   \int\frac{d^{3}K}{\sqrt{E_{Q}E_{S}}}\Psi_{\Lambda_{Q}}(\vec{K})\mid
   \vec{q} (\vec{P}-\vec{K}),\lambda;S(\vec{K})>,
\end{equation}		

\noindent where color indices have been omitted, $E_{Q}$ and $E_{S}$ are 
the IMF energies of the
heavy quark and scalar-particle, respectively, $\lambda$ represses 
the helicity of
the baryon. State normalization is taken as

\begin{equation}
<\Lambda_{Q}(\vec{P'}),{\lambda}'\mid\Lambda_{Q},(\vec{P}),{\lambda}>
  = \frac{E_{\Lambda_{Q}}}{m_{\Lambda_{Q}}}\delta(\vec{P}-\vec{P'})
    \delta_{\lambda{'}\lambda},
\end{equation}
which results in the following normalization of the baryon wave function
$\Psi_{\Lambda_{Q}}(x_{1},\vec{K}_{\perp})$

\begin{equation}
\int{dx_{1}}d^{2}K_{\perp}\mid\Psi_{\Lambda_{Q}}(x_{1},\vec{K}_{\perp})\mid^{2} = 1. 
\end{equation}

\noindent Here, the longitudinal momentum 
fraction $x_{1}$ carried by the heavy quark and
the heavy quark's transverse momentum corresponding to its parent baryon
$\vec{K}_{\perp}$ are introduced. Obviously, the scalar-particle[ud] 
carries $x_{2}=1-x_{1}$
and -$\vec{K}_{\perp}$. The baryon wave function $\Psi_{\Lambda_{Q}}(x_{1},\vec{K}_{\perp})$
is a generalization of the BSW$^{[10]}$ meson
wave function to the quark-diquark case

\begin{equation}
\Psi_{\Lambda_{Q}}(x_{1},\vec{K}_{\perp}) = N_{\Lambda_{Q}}x_{1}x_{2}^3exp[-b^{2}(\vec{K}_{\perp}^2
   + m_{\Lambda_{Q}}^{2}(x_{1}-x_{0})^{2})].
\end{equation}

\noindent The peak position of the wave function 
is at $x_{0}=1-\varepsilon/m_{\Lambda_{Q}}$, 
where the parameter $\varepsilon$ is the difference 
between the hadron and the heavy-quark (constituent) mass
and has a value of about 0.6 GeV. This is almost the constituent mass 
of the diquark. Another parameter b in the wave function
is related to the mean $K_{\perp}$ or the radius of 
the baryon and its precise value
is not known. However, we expect the radius of a heavy baryon
to be smaller than that of proton.
In the following calculations, as in ref. [8], we use b=1.77GeV and b=1.18
GeV, corresponding to $<K_{\perp}^{2}>$$^{\frac{1}{2}}>$ = 400 MeV and 
$<K_{\perp}^{2}>$$^{\frac{1}{2}}>$ = 600MeV respectively. The wave
function overlap integral for $F_{1}$ and $F_{2}$ can be easily 
obtained by the matrix elements of the so-call good current components 
($\mu=0,3$)

\begin{equation}
F_{1} + F_{2} = C_{s} I(v),
\end{equation}

\noindent where $I(v)$ is the overlap integral      
		
\begin{equation}
I(v) = \sqrt{\frac{m_{\Lambda_{b}}}{m_{\Lambda}}}
   \int_{1-\frac{m_{\Lambda}}{m_{\Lambda_{b}}}}^{1}dx
   \int_{-\infty}^{\infty}d^{2}k_{\perp}
   \Psi_{\Lambda}(1-\frac{m_{\Lambda_{b}}}{m_{\Lambda}}(1-x),
   \vec{K}_{\perp}+(1-x)m_{\Lambda_{b}}v\vec{e}_{1})
   \Psi_{\Lambda_{b}}(x,\vec{K}_{\perp}).
\end{equation}

\noindent Here $\vec{e}_{1}$ represents the unit vector in $x$ direction,
the occurrence of the parameter $C_{s}$ is because that $\Lambda$ has 
to be considered
as a superposition of various quark-diquark configuration$^{[12]}$ but 
can not be 
regarded as being made just of a strang quark and a quasi-particle$[ud]$.
 However,
in our case, only the $sS[ud]$ state can contribute 
to $\Lambda_{b}\rightarrow\Lambda$ decay and thus
the overlap integral is suppressed by an appropriate Clebsch-Gordan coefficient
C$_{s}$ which is 1/$\sqrt{3}$ in the model of [12]. Replacing 
the argument $v$ in $I$ by the 
invariant momentum transfer q$^2 = (P_{\Lambda_{b}}-P_{\Lambda})^{2}$, 
from (8) and (16), we get

\begin{equation}
F_{1} = \frac{2E_{\Lambda}+m_{\Lambda}+m_{s}}{2(E_{\Lambda}+m_{s})}
     C_{s}I(q^{2}),
\end{equation}

\begin{equation}
F_{2} = \frac{m_{s}-m_{\Lambda}}{2(E_{\Lambda}+m_{s})}
     C_{s}I(q^{2}).
\end{equation}

\noindent The form factors $F_{1}$ and $F_{2}$   are
plotted in Fig.1 as functions of 
$\omega(=v_{\Lambda_{b}}\cdot{P_{\Lambda}/m_{\Lambda}}$).

\section{ Branching Ratio for $\Lambda_{b}\rightarrow\Lambda + J$/$\psi$}

\hspace{0.29cm} In this section, we will discuss the process
$\Lambda_{b}\rightarrow\Lambda + J$/$\psi$ and calculate its 
branching ratio. This process proceeds only through
the internal W-emission diagram and under factorization assumption its weak
decay amplitude reads

\begin{equation}
A(\Lambda_{b}\rightarrow\Lambda + J/\psi) = \frac{G_{F}}{\sqrt{2}}V_{cb}
  V_{cs}^{*}a_{2}<J/{\psi}\mid\bar{c}\gamma_{\mu}(1-\gamma_{5})c\mid{0}>
  <\Lambda\mid\bar{s}\gamma_{\mu}(1-\gamma_{5})b\mid\Lambda_{b}>,
\end{equation}
\noindent where $a_{2}$ is a free parameter necessary to be determined 
experimentally, $V_{cb}$ and $V_{cs}$
are CKM matrix elements and G$_{F}$ is Fermi coupling constant. 
The matrix element
of $\Lambda_{b}\rightarrow\Lambda$ can be generally defined as
the following on the ground of Lorentz decomposition
\begin{eqnarray}
 <\Lambda(P_{\Lambda})\mid\bar{s}\gamma_{\mu}(1-\gamma_{5})b\mid
 \Lambda_{b}(P_{\Lambda_{b}})>&=
&\bar{U}_{\Lambda}[f_{1}(q^{2})\gamma_{\mu} + {i}f_{2}(q^{2})\sigma
 _{\mu\nu}q^{\nu} + f_{3}(q^{2})q_{\mu}-(g_{1}(q^{2})\gamma_{\mu} \nonumber\\
 &   & + {i}g_{2}(q^{2}) 
 \sigma_{\mu\nu}q^{\nu} + g_{3}(q^{2})q_{\mu})\gamma_{5}]U_{\Lambda_{b}},
\end{eqnarray}

\noindent where $f_{i}$ and $g_{i}$ are related to $F_1$ and $F_2$ by

\begin{equation}
f_{1}(q^{2}) = g_{1}(q^{2}) = F_{1}(q^{2}) + \frac{m_{\Lambda}}
   {m_{\Lambda_{b}}}F_{2}(q^{2}) ,
\end{equation}

\begin{equation}
f_{2}(q^{2}) = g_{2}(q^{2}) = f_{3}(q^{2}) =
   g_{3}(q^{2}) = \frac{1}{m_{\Lambda_{b}}}F_{2}(q^{2}).
\end{equation}
					
\noindent Comparing with the general amplitude of 
$\Lambda_{b}\rightarrow\Lambda + J$/$\psi$                

\begin{equation}
A(\Lambda_{b}\rightarrow\Lambda + J/\psi) = i \bar{U}_{\Lambda}
 (P_{\Lambda})\varepsilon^{*\mu}[A_{1}\gamma_{\mu}\gamma_{5}
 + A_{2}(P_{\Lambda})_{\mu}\gamma_{5} + B_{1}\gamma_{\mu}
 + B_{2}(P_{\Lambda})_{\mu}]U_{\Lambda_{b}}(P_{\Lambda_{b}}),
\end{equation}

\noindent and using (21), (22) and (23) lead to

\begin{equation}
A_{1} = -\eta[F_{1}(m_{J/\psi}^{2}) + F_{2}(m_{J/\psi}^{2})],
\end{equation}

\begin{equation}
A_{2} = -2\eta\frac{1}{m_{\Lambda_{b}}}F_{2}(m_{J/\psi}^{2}),
\end{equation}

\begin{equation}
B_{1} = \eta[F_{1}(m_{J/\psi}^{2}) - F_{2}(m_{J/\psi}^{2})],
\end{equation}

\begin{equation}
B_{2} = 2\eta\frac{1}{m_{\Lambda_{b}}}F_{2}(m_{J\psi}^{2}),
\end{equation}

\noindent with $\eta =\frac{G_{F}}{\sqrt{2}}V_{cb}
  V_{cs}^{*}a_{2}f_{J/\psi}m_{J/\psi}$, where $f_{J/\psi}$ is the $J/\psi$ 
decay constant and $m_{J/\psi}$ expresses the mass of $J/\psi$.  
The decay width is given by $^{[13]}$

\begin{equation}
 \Gamma(\Lambda_{b}\rightarrow\Lambda + J/\psi) = \frac{1}{8\pi}
 \frac{E_{\Lambda} + m_{\Lambda}}{m_{\Lambda_{b}}}P_{J/\psi}
 [2(\mid{S}\mid^{2} + \mid{P_{2}}\mid^{2}) + \frac{E_{J/\psi}^{2}}
 {m_{J/\psi}^{2}}(\mid{S + D}\mid^{2} + \mid{P_{1}}\mid^{2})].
\end{equation}

\noindent Here, $P_{J/\psi}$ and $E_{J/\psi}$  are the momentum and energy 
of $J/\psi$ in the rest frame of $\Lambda_{b}$
respectively and 

\begin{equation}
 S = -A_{1},
\end{equation}

\begin{equation}
 D = -\frac{P_{J/\psi}^{2}}{E_{J/\psi}(E_{\Lambda}+m_{\Lambda})}
   (A_{1} - m_{\Lambda_{b}}A_{2}),
\end{equation}

\begin{equation}
 P_{1} = -\frac{P_{J/\psi}}{E_{J/\psi}}(\frac{m_{\Lambda_{b}} + m_{\Lambda}}
   {E_{\Lambda} + m_{\Lambda}}B_{1} + m_{\Lambda_{b}}B_{2}),
\end{equation}

\begin{equation}
 P_{2} = \frac{P_{J/\psi}}{E_{\Lambda}+m_{\Lambda}}B_{1}.
\end{equation}

Using the numerical values of $F_{1}$ and $F_{2}$ at $m_{J\psi}$ in the limit 
$m_{b}\rightarrow\infty$, we obtain the
width and branching ratio of $\Lambda_{b}\rightarrow\Lambda + J$/$\psi$     

\begin{eqnarray}
\Gamma(\Lambda_{b}\rightarrow\Lambda + J/\psi) = \left  \{ 
            \begin{array}{ll}
            1.83\times10^{-17}GeV  & \mbox{ when $b=1.18GeV^{-1}$,}\\
            1.19\times10^{-18}GeV  & \mbox{ when $b=1.77GeV^{-1}$.}
            \end{array}
                                                 \right.     
\end{eqnarray}

\begin{eqnarray}
BR(\Lambda_{b}\rightarrow\Lambda + J/\psi) = \left \{
                   \begin{array}{ll}
            2.97\times10^{-5}  &  \mbox{ when $b=1.18GeV^{-1}$,}\\
            1.94\times10^{-5}  &  \mbox{ when $b=1.77GeV^{-1}$.}
                   \end{array}
                                            \right.    
\end{eqnarray}
		
\noindent The up-down asymmetry parameter $\alpha$ given by$^{[10]}$

\begin{equation}
\alpha = \frac{4m_{J/\psi}^{2}Re(S^{*}P_{2})+2E_{J/\psi}^{2}
  Re(S+D)^{*}P_{1}}{2(\mid{S}\mid^{2}+\mid{P_{2}}\mid^{2})m_{J/\psi}^{2}
  + (\mid{S+D}\mid^{2} + \mid{P_{1}}\mid^{2})E_{J/\psi}^{2}},
\end{equation}

\noindent is numerically found to be 

\begin{equation}
  \alpha(\Lambda_{b}\rightarrow\Lambda + J/\psi)
    = -0.19.
\end{equation}

\par
Some of parameters used in calculations are chosen as: $V_{cb}=0.04$,
$V_{cs}=0.97$, $f_{J/\psi}=0.395GeV$, $m_{\Lambda_{b}} = 5.64GeV$,
$m_{\Lambda} = 1.116GeV$, $m_{s} = 0.15GeV$, $\varepsilon = 0.6GeV$, 
$\tau(\Lambda_{b}) = 1.07\times10^{-12}s$.\par
$a_2$ in eq. (20) has some uncertainty. In principle it is related to 
hadronization and at present it can only be determined by experiment. There
are some discussions on it [14]. In the above calculation we choose $a_2
=0.23$ [7]. 

\section {Conclusion}

\hspace{0.29cm} To sum up, to study the transition 
$\Lambda_{b}\rightarrow\Lambda$, the two dynamical
assumptions suggested by Stech in the meson case are generalized to the 
baryon case. This leads to a relation between the two form factors
$F_{1}$ and $F_{2}$ in heavy quark limit. Further more, to determine
$F_{1}$ and $F_{2}$ absolutely, 
we apply the model of Guo and Kroll to our case.
Making use of the form factors $F_1$ and $F_2$ obtained, 
the width of decay and branching ratio of $\Lambda_{b}\rightarrow\Lambda + J$/$\psi$
are calculated. In spite of the sensitivities 
of the width of decay
and branching ratio for $\Lambda_{b}\rightarrow\Lambda + J$/$\psi$ to 
the parameter b, which reflects the soft
dynamics in the weak transition, we conclude that 
BR($\Lambda_{b}\rightarrow\Lambda + J$/$\psi$) is of 
the order of $10^{-5}$, which is the same as that obtained by
Datta but smaller than that in ref. [7]. This one order difference may arise from the assumption of the flavor independence of hadronic wave functions in ref. [7].   
The up-down asymmetry parameter $\alpha$ is equal to -0.19 and basically
in accordance with that arrived at in [7]. It is
noted that the parameter $\alpha$ in our approach has nothing to do 
with the overlap integral of
wave function and depends only on the Stech's dynamical assumptions 
generalized to baryon case.
In other words, if the spectator's spin and momentum remain unchanged at the first stage of interaction and if its off-shell energy is almost a constant in the rest frame of its parent baryon this parameter is determined.
In the present work , we proceeded only in the heavy quark limit    
and thus a correction of $1/m_{b}$ is necessary to improve our results.
However, because of large mass of b-quark the results including $1/m_{b}$ corrections
will not improve much over the present results.\par  
\vspace{1cm}

\noindent Acknowledgment:

One of us (Guo) would like to thank TWAS and CBPF for the finacial support. Part of the work is done in CBPF. This work is in part supported by the National Science Foundation of China.

\newpage

\noindent{\bf Figure Captions}

\vspace{1cm}
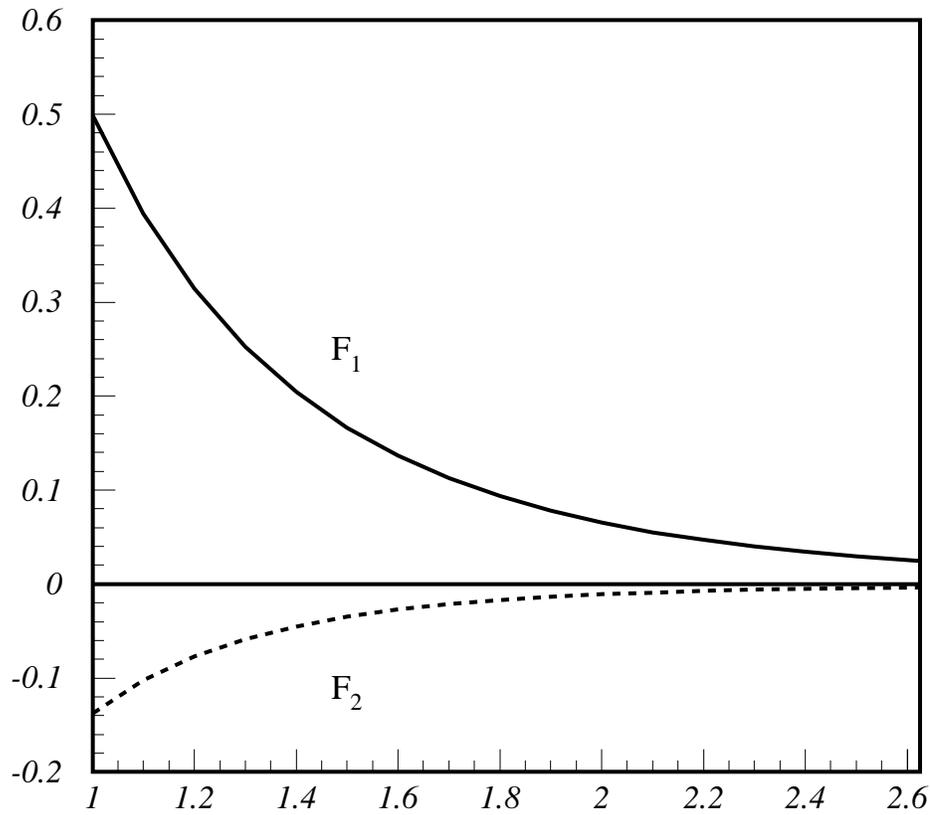
\noindent Fig. 1 $\Lambda_{b}\rightarrow\Lambda$ form factors $F_{1}$ and $F_{2}$ 
corresponding to $\omega(=v_{\Lambda_{b}}\cdot{P_{\Lambda}/m_{\Lambda}}) (b=1.77GeV^{-1})$.

\end{document}